# Parallel Markov Chain Monte Carlo for Non-Gaussian Posterior Distributions

Alexey Miroshnikov[1], Erin M. Conlon[1*]

[1] Department of Mathematics and Statistics, University of Massachusetts, Amherst, Massachusetts, United States of America

* Corresponding author

E-mail: econlon@mathstat.umass.edu

**Abstract**

Recent developments in big data and analytics research have produced an abundance of large data sets that are too big to be analyzed in their entirety, due to limits on computer memory or storage capacity. To address these issues, communication-free parallel Markov chain Monte Carlo (MCMC) methods have been developed for Bayesian analysis of big data. These methods partition data into manageable subsets, perform independent Bayesian MCMC analysis on each subset, and combine the subset posterior samples to estimate the full data posterior. Current approaches to combining subset posterior samples include sample averaging, weighted averaging, and kernel smoothing techniques. Although these methods work well for Gaussian posteriors, they are not well-suited to non-Gaussian posterior distributions. Here, we develop a new direct density product method for combining subset marginal posterior samples to estimate full data marginal posterior densities. Using a commonly-implemented distance metric, we show in simulation studies of Bayesian models with non-Gaussian posteriors that our method outperforms the existing methods in approximating the full data marginal posteriors. Since our method estimates only marginal densities, there is no limitation on the number of model parameters analyzed. Our procedure is suitable for Bayesian models with unknown parameters with fixed dimension in continuous parameter spaces.

# Introduction

Due to the exponential growth of big data and analytics in recent years, statisticians are facing new challenges in the analysis of large data sets. Here, big data is defined as data sets that are too large and complex for traditional analysis tools to be used. The application areas affected by big data include genomics, sustainability, healthcare, finance, energy and meteorology, among many others. One main difficulty in the analysis of large data sets is that they are too big to be analyzed in their entirety, due to limits on either computer memory or storage capacity; in addition, the processing time may be excessive for complete data sets. To address these issues, several recent Bayesian and Markov chain Monte Carlo (MCMC) methods for big data have been developed. One approach divides big data sets into smaller subsets, and analyzes the subsets on separate machines using parallel MCMC computation (Langford et al. [1], Newman et al. [2], Smola and Narayanamurthy [3]); here, communication between machines is required, since information is exchanged for each sample of the Markov chains.

Due to the slow performance of the communication-based methods, new techniques have been derived that do not require communication between machines. One research direction involves copying the full data set onto each machine and performing independent, parallel MCMC on each machine (Wilkinson [4], Laskey and Myers [5], Murray [6]). However, these methods are not appropriate when the full data set is either too large to be read into computer memory or when storage capacity on each machine is limited. Another research avenue involves partitioning the data into subsets, implementing independent Bayesian MCMC computation on the subsets, and joining the independent results back together (Neiswanger et al. [7], Scott et al. [8]). These are parallel, communication-free methods, i.e. embarrassingly parallel methods, in that each machine (or subset) creates MCMC samples without communicating with any other machine. These algorithms are particularly well-suited to the MapReduce framework (Dean and Ghemawat [9]), and can be run on parallel computing systems such as Hadoop (White [10]), on networks of machines, and on multi-core processors. In addition, existing Bayesian software programs can be used to produce the subset posterior samples, such as WinBUGS [11], JAGS [12] and Stan [13,14].

The methods of Neiswanger et al. [7] and Scott et al. [8] have different strategies for combining subset posterior samples. Specifically, Neiswanger et al. [7] introduced a kernel density estimator that first approximates each subset posterior density; the full data posterior is then estimated by multiplying the subset posterior densities together. These authors produced an algorithm that generates samples from the distribution that approximates the full data kernel density estimator through a form of an MCMC sampler. Alternatively, Scott et al. [8] developed the consensus Monte Carlo algorithm that uses weighted averages over the subset MCMC samples to estimate the full data posterior. The methods of Neiswanger et al. [7] and Scott et al. [8] perform well in practice for subset posteriors that are near Gaussian, which is expected when the subset sample size is adequately large, due to the Bayesian central limit theorem (Bernstein von-Mises theorem; see Van der Vaart [15] and Le Cam and Yang [16]). Yet, for non-Gaussian posteriors, these methods do not perform as well. In Neiswanger et al. [7] it is stated that the algorithm is asymptotically exact, regardless of whether the posterior is Gaussian or non-Gaussian. However, the authors point out that for a finite number of MCMC samples, this method produces near-Gaussian posteriors, regardless of the shape of the true underlying full data distribution. The method of Neiswanger et al. [7] also has difficulty when the number of unknown model parameters is large, e.g. greater than 50, since it becomes infeasible to approximate joint densities through kernel density estimation as the number of parameters increases.

Here, we introduce a new embarrassingly parallel MCMC algorithm named the direct density product method that performs well for non-Gaussian posterior distributions, and is not limited by the number of unknown model parameters. Our general approach is in the spirit of the work of Neiswanger et al. [7] in that we estimate the full data posterior density using the product of subset posterior densities. However, our method for carrying out the estimation differs greatly from Neiswanger et al. [7]; we detail these differences in the Methods section below. Our method produces only estimated marginal distributions rather than joint distributions, and is thus especially useful for models in large dimensions. In contrast, the methods of Neiswanger et al. [7] and Scott et al. [8] produce estimated joint posteriors and thus have more computational difficulty as the model dimension grows. Note that there are other methods in Bayesian settings that focus on estimating only marginal posteriors, including the Integrated Nested Laplace Approximation (INLA) method of Rue et al. [17].

In addition to the above methods for big data analysis in a Bayesian framework, alternative approaches have been developed, but each has limitations. Huang and Gelman [18] introduced importance resampling strategies, but these methods have difficulties in that they can collapse to a single point when the parameter space has high dimension. Other methods include INLA for large data sets (Rue et al. [17]), but this technique has computational cost that increases exponentially with the number of unknown model parameters.

Here, we implement our direct density product method using several Bayesian models that produce non-Gaussian posterior distributions. Using a commonly-implemented metric, we show that our method outperforms the methods of Neiswanger et al. [7], Scott et al. [8], and the third method of simple averaging of subset samples.

Our paper is organized as follows. In the Methods section, we introduce our new direct density product method and describe the three comparison methods for combining subset posterior samples. In the following section, we introduce several Bayesian models that generate non-Gaussian posteriors, including single parameter and multiparameter models, and compare results of the four methods. We summarize our findings in the Discussion section.

**Methods**

For Bayesian models, the posterior distribution given the full data set is the following, for the vector of unknown model parameters $\theta = (\theta^1, \theta^2, ..., \theta^d) \in \mathbb{R}^d$ of dimension $d \geq 1$,

$$p(\theta | y) \propto p(y | \theta) p(\theta). \tag{1}$$

Here, $p(\mathbf{y}|\theta)$ is the likelihood of the full data set given $\theta$, and $p(\theta)$ is the prior distribution of $\theta$. For big data sets, $\mathbf{y}$ is assumed to be too large to analyze entirely, and $\mathbf{y}$ is thus randomly partitioned into $M$ disjoint subsets (or machines) $\mathbf{y}_m$, $m=1,...,M$. The data set $\mathbf{y}$ is partitioned by the $r$ data values, so that if $\mathbf{y}$ has dimension $r \times s$, then $\mathbf{y}$ is partitioned as follows:

$$\mathbf{y} = \begin{pmatrix} \mathbf{y}_1 \\ \mathbf{y}_2 \\ \vdots \\ \mathbf{y}_m \end{pmatrix}, \qquad (2)$$

where each $\mathbf{y}_m, m=1,...,M$, has $s$ columns. We sample from each posterior density of $\theta$ given each data subset $\mathbf{y}_m$, defined as: $p_m(\theta|\mathbf{y}_m)$, $m=1,...,M$; these are labeled as subposterior samples. Assuming independence of the subposterior densities, the samples from the subposterior densities are combined to estimate the posterior density given the full data set, using the following expression:

$$p(\theta|\mathbf{y}) \propto p(\mathbf{y}|\theta)p(\theta) \propto \prod_{m=1}^{M} p_m(\theta|\mathbf{y}_m) \propto \prod_{m=1}^{M} p(\mathbf{y}_m|\theta)p(\theta)^{1/M}. \qquad (3)$$

Here, the prior distribution $p(\theta) = \prod_{m=1}^{M} p(\theta)^{1/M}$, so that the total amount of prior information is equivalent in the full-data model and the independent subposterior density product model.

In the following sections, we describe each of the four methods for combining the independent subposterior samples, including simple averaging across subset samples, the methods of Neiswanger et al. [7] and Scott et al. [8], and our new direct density product method. We denote the subposterior samples as $\theta_{t,m} = \left(\theta_{t,m}^1, \theta_{t,m}^2, ..., \theta_{t,m}^d\right)$, for $d$ unknown model parameters, subset $m$, $m = 1,…,M$, and MCMC iteration $t$, $t = 1,…,T$; these samples have been drawn from each of the subposterior densities

$$p_m(\theta) \propto p(\mathbf{y}_m | \theta) p(\theta)^{1/M}, \quad m = 1,...,M.$$

**Average of subposterior samples method**

For subposterior sample $\theta_{t,m}$, for MCMC iteration $t$, $t = 1,…,T$, and subset $m$, $m = 1,…,M$, the independent subposterior samples are pooled into the combined posterior samples $\theta_t$, $t = 1,...,T$, by averaging the subposterior samples over the subsets for each MCMC iteration $t$, as follows:

$$\theta_t = \frac{1}{M}\sum_{m=1}^{M} \theta_{t,m}, \quad t = 1,...,T. \tag{4}$$

Here, the $d$ individual unknown model parameters are assumed to be independent.

**Consensus Monte Carlo method**

The consensus Monte Carlo method was developed by Scott et al. [8]. This method pools the independent subposterior samples across subsets within an MCMC iteration into the combined joint posterior samples $\theta_t$, $t = 1,...,T$, using weighted averages, as follows:

$$\theta_t = \left(\sum_{m=1}^{M} W_m\right)^{-1} \left(\sum_{m=1}^{M} W_m \theta_{t,m}\right), \quad t = 1,...,T. \tag{5}$$

Here, $W_m = \Sigma_m^{-1}$ for each subset $m$, $m = 1,…,M$, where $\Sigma_m = \text{Var}(\theta | \mathbf{y}_m)$ is the $d \times d$ - dimensional variance-covariance matrix for the $d$ unknown model parameters. For this

method, $\Sigma_m$ is estimated by the sample variance-covariance matrix based on the $T$ MCMC subposterior samples $\theta_{t,m}$, $t = 1,...,T$.

**Semiparametric density product estimator method**

The next approach uses kernel density estimators, and was introduced by Neiswanger et al. [7]. Here, each subposterior density is estimated using a Gaussian kernel to smooth the subset MCMC subposterior samples. The product of the subposterior densities approximates the full data posterior density; this product consists of a mixture of $T^M$ Gaussians with corresponding mixture weights, for $T$ MCMC samples and $M$ subsets. Samples are generated from this mixture by first randomly sampling one of the $T^M$ Gaussian mixture components, and then sampling from the selected component. Results of this method produce an asymptotically unbiased estimate of the full data joint posterior density. The algorithm was named the semiparametric density product estimator (DPE) method by Neiswanger et al. [7] and it generates the combined joint posterior samples $\theta_t$, $t = 1,...,T$.

**Direct density product method**

In our new direct density product method, we first estimate the independent subset marginal posterior densities $p_m^j(x)$ of the parameter $\theta^j$, $j = 1,...,d$ (to simplify notation, we use $j$ in place of $\theta^j$ throughout; e.g., we use $p_m^j(x)$ rather than $p_m^{\theta^j}(x)$). The product of the estimated independent subset posterior densities then approximates the full data marginal posterior density $p^{j,PRE}(x)$, for the parameter $\theta^j$, as follows:

$$\hat{p}^{j,PRE}(x) = \widehat{p_1^j \cdots p_M^j}(x) = \hat{p}_1^j \cdots \hat{p}_M^j(x) = \prod_{m=1}^{M} \hat{p}_m^j(x); \tag{6}$$

we use the superscript "*PRE*" for preliminary, since the product in Equation (6) produces, in general, an unnormalized density estimate. To normalize this preliminary density estimate, we approximate it by polynomials and use this approximation to compute the normalization constant; this procedure is described next. Specifically, our method consists of the following four steps for each individual unknown model parameter $\theta^j$, $j = 1, ..., d$:

1) Estimate each marginal subset posterior density, $p_m^j(x)$, using density estimation. For this step, any density estimation technique can be used, including nonparametric density estimation methods such as kernel density estimation (see Silverman [19], Scott [20], Rosenblatt [21] and Parzen [22]) or penalized likelihood approaches (see Silverman [19], Scott [20] and Schellhase and Kauermann [23]). For the density estimation, we specify a fixed grid of values on the *x*-axis: $a = x_0 < x_1 < ... < x_n = b$. Here, the range [a, b] is chosen so that it contains the range of the MCMC samples for each parameter $\theta^j$; extra width extends beyond the range of the MCMC samples to allow the density to reduce to approximately zero at the extremes. The density estimation then generates the series of points $(x_i, \hat{p}_m^j(x_i))$, $i = 0, ..., n$, where $\hat{p}_m^j(x_i)$ is the estimated density evaluated at each $x_i$.

2) We evaluate the unnormalized full data density estimate $\hat{p}^{j,PRE}(x)$ at each grid value $x_i$ by multiplying all estimated subset marginal density values, as follows:

$$\hat{p}^{j,PRE}(x_i) = \prod_{m=1}^{M} \hat{p}_m^j(x_i), \quad i = 0, ..., n. \tag{7}$$

The next two steps are carried out in order to normalize this density to integrate to 1.

3) The previous step 2) results in a series of points $(x_i, \hat{p}^{j,PRE}(x_i))$, $i = 0,...,n$. Next, we produce an estimated density curve through these points using standard Lagrange polynomial interpolation of order $k \geq 1$ (see Atkinson [24]). This produces the interpolated density estimate of $\hat{p}^{j,PRE}(x)$, denoted by $\hat{\hat{p}}^{j,PRE}(x)$, which is a polynomial of order $k$ on each subinterval $[x_{rk}, x_{(r+1)k}]$, $r = 0,..., N-1$, where $n = kN$. Note that at the grid values $x_i$, $i = 0,...n$, we have $\hat{\hat{p}}^{j,PRE}(x_i) = \hat{p}^{j,PRE}(x_i)$.

4) We normalize the resulting density in step 3) so that it integrates to 1. For this, $\hat{\hat{p}}^{j,PRE}(x)$ is integrated to produce the estimated area $c$ under the curve; this value $c$ is used as a normalizing constant. We use the standard composite Newton-Cotes integration formulas to calculate $c$ (Atkinson [24]). The result of this step is the function $\dfrac{\hat{\hat{p}}^{j,PRE}(x)}{c}$, defined on [a,b], which approximates the full data posterior at any value of $x$ on [a,b].

Note that since $\hat{p}^{j,PRE}(x_i) = \hat{\hat{p}}^{j,PRE}(x_i)$, $i = 0,...n$, our method generates the series of points: $\left(x_i, \dfrac{\hat{p}^{j,PRE}(x_i)}{c}\right)$, $i = 0,...,n$; we denote these as the normalized values $\left(x_i, \hat{p}^j(x_i)\right)$, $i = 0,...,n$.

If a user wishes to produce samples from each estimated posterior marginal density in step 4), these can be generated using the standard inverse cumulative distribution function sampling method; we provide the standard algorithm for this in the Appendix.

Our method is similar in essence to the semiparametric DPE method, only in that it uses density estimation for each subset posterior density. However, the semiparametric DPE method uses kernel density estimation specifically, while our method can use any density estimation method. It is known that kernel density estimators have difficulty in estimating non-smooth densities (see Van Eeden [25]). In addition, the semiparametric DPE method can require sampling from a mixture of $5 \times 10^{80}$ Gaussian components when $T = 50,000$ MCMC samples and $M = 20$ subsets (for example), which makes accurate sampling both difficult and time consuming. In contrast, our method only requires multiplication of the estimated subset density values at each *x*-axis grid value. Note again that the semiparametric DPE method produces joint posterior density estimates, while our new method produces only estimated marginal posterior densities. Thus, our method is computationally feasible even for high-dimensional models.

**A metric for comparing densities**

Results for the four methods described above are compared using an estimate of the $L_2$ distance, $d_2(p, \hat{p})$, between the marginal posterior density $p$ based on the full data set and the estimated marginal posterior $\hat{p}$. This value was introduced in Neiswanger et al. [7], and is defined by the following, for parameter $\theta^j$, $j = 1,...,d$:

$$d_2(p,\hat{p}) = \|p - \hat{p}\|_{L_2} = \left( \int_{\mathbb{R}} \left( p(\theta^j) - \hat{p}(\theta^j) \right)^2 d\theta^j \right)^{1/2}. \tag{8}$$

Here, we calculate $L_2$ distance using density smoothing for both $p$ and $\hat{p}$, as described in Neiswanger et al. [7] and Oliva et al. [26]. The estimated relative $L_2$ distance, relative to the full data marginal posterior, is produced for each of the four methods in our examples in the following section. This value is calculated for each marginal posterior density, and averaged over all model parameters.

## Examples with non-Gaussian posterior distributions

Here, we illustrate our new direct density product method through simulation studies of several Bayesian statistical models with non-Gaussian posterior distributions. The performance of our new method is compared to the three methods described above, using estimated relative $L_2$ distances. For each of the following examples, data sets of size 100,000 are simulated, so that full data analyses are still feasible. We sampled 50,000 MCMC iterations after burnin of 5,000 iterations for each of the unknown model parameters, for both the subset analyses and the full data analyses, using either the R programming language [27] or WinBUGS [11]. The number of subsets was set to $M = 10$ and $M = 20$ in each example. For density estimation, we use the penalized likelihood logspline method, which uses splines to estimate the log-density based on the subset MCMC samples (Kooperberg and Stone [28,29] and Stone and Koo [30]). This method is implemented using the R package **logspline** (Kooperberg [31]). We use this density estimation method since it performs well for bounded posteriors, which are encountered in our models. In the examples presented below, we use evenly-spaced values for the *x*- axis grid, with sub-interval width of $\Delta x_i = x_{i+1} - x_i = 10^{-5}$, $i = 0,\ldots,n$-1; we also use quadratic polynomials for the density curve interpolation, i.e., $k = 2$.

**Binomial distribution with conjugate prior**

Our first model is a Bernoulli model with unknown success probability $p$. We specify $p$ = 0.001, corresponding to a rare event; a small value of $p$ is chosen so that the posterior of $p$ has a skewed, non-Gaussian distribution. A conjugate Beta prior is assigned for $p$, which has the following form:

$$p(p|\alpha,\beta) \propto p^{\alpha-1}(1-p)^{\beta-1}. \qquad (9)$$

For the prior distribution for the full data model, we assign prior parameters $\alpha = 1, \beta = 1$, which is equivalent to a Uniform(0,1) distribution. For the subset priors, we require the product of these priors to be proportional to the full data prior, as follows:

$$\prod_{m=1}^{M} p^{\alpha'-1}(1-p)^{\beta'-1} \propto p^{\alpha-1}(1-p)^{\beta-1}. \qquad (10)$$

We thus need the following equivalence, for $M$ subsets:

$$M(\alpha'-1) = \alpha - 1, \qquad (11)$$

$$M(\beta'-1) = \beta - 1, \qquad (12)$$

where $\alpha'$ and $\beta'$ are the subset prior parameters. We solve for $\alpha'$ and $\beta'$, from which it follows that:

$$\alpha' = \frac{\alpha + M - 1}{M}, \qquad (13)$$

$$\beta' = \frac{\beta + M - 1}{M}. \qquad (14)$$

For $M = 10$ subsets and $\alpha = 1$ and $\beta = 1$ for the full data prior, the resulting values are $\alpha' = 1$ and $\beta' = 1$. For $M = 20$ subsets, the resulting subset prior parameters are also $\alpha' = 1$ and $\beta' = 1$. Note that, for the subset data, many subsets contain zero events.

After posterior MCMC sampling for the $M = 10$ data subsets as well as the full data set, we found that the direct density product method has the lowest estimated relative $L_2$ distance of the four methods, with a value of 0.015. In contrast, the remaining three methods have considerably larger estimated relative $L_2$ distances, ranging from 0.288 to 0.576 (Table 1), which indicates that these methods have limitations when estimating non-Gaussian posterior distributions. We plot the estimated combined posterior density for $p$ for each of the four combining methods in Figure 1; we also plot the full data posterior and the ten subset posterior densities. When increasing the number of subsets to $M = 20$ while the data size remains constant, the estimated relative $L_2$ distance increases for all four methods. This is due to the decrease in data sample size per subset as the number of subsets increases. The estimated relative $L_2$ distance for the direct density product method is still reasonably low, with a value of 0.054, while the remaining three methods have appreciably larger estimated relative $L_2$ distances ranging from 0.510 to 1.015 (Table 1).

**Multinomial distribution with conjugate prior**

Our next example is a Multinomial model with unknown probability vector $p$, with a conjugate Dirichlet prior for $p$; here, $p$ has dimension $d = 20$. We simulate data from the Multinomial distribution with extreme event probabilities $= 0.001$ for all but the final category, which has event probability = 0.981, so that the probabilities sum to 1. These values for $p_j$, $j = 1,\ldots,20$, are chosen so that the posteriors are again skewed, non-Gaussian distributions. The Dirichlet prior for $\boldsymbol{p}$ has the following form:

$$p(\boldsymbol{p}|\boldsymbol{\alpha}) \propto \prod_{j=1}^{20} p_j^{\alpha_j - 1}. \tag{15}$$

For the prior for the full data model, we assign the prior parameters $\alpha_j = 1$, $j = 1,\ldots,20$; this corresponds to a Uniform prior for $\boldsymbol{p}$ over the open standard $(d-1)$-dimensional simplex. For the $M$ subset priors, we again require the product of the subset priors to be proportional to the prior for the full data model, as follows:

$$\prod_{m=1}^{M} \left( \prod_{j=1}^{20} p_j^{\alpha'_j - 1} \right) \propto \prod_{j=1}^{20} p_j^{\alpha_j - 1}. \tag{16}$$

This forms the following equivalence:

$$M\left(\alpha'_j - 1\right) = \alpha_j - 1, \quad j = 1,\ldots,20, \tag{17}$$

where $\alpha'_j$ are the subset prior parameters. Solving for $\alpha'_j$ produces the following:

$$\alpha'_j = \frac{\alpha_j + M - 1}{M}, \quad j = 1,\ldots,20. \tag{18}$$

This results in $\alpha'_j = 1$, $j = 1,\ldots,20$, for the subset priors, for both $M = 10$ and $M = 20$.

We then produced the MCMC samples for the $M = 10$ data subsets as well as for the full data set. Our results show that the direct density product method has the lowest average estimated relative $L_2$ distance of the four methods for the marginal posterior distributions of the $p_j$, $j = 1,\ldots,19$ (Table 1). Note that we only estimate the first 19 marginal posterior distributions, since the value of the 20th parameter is determined by the first 19, since the sum of all parameters is 1. The average was 0.029 for the direct density product method, while the values were markedly larger for the remaining three methods, ranging from 0.194 to 0.581 (Table 1). This again illustrates that the three comparison methods have some difficulties when estimating non-Gaussian posterior distributions. Figure 2 displays the estimated combined posterior density and full data posterior density for the marginal of $p_1$ for all four methods; we also show the full data posterior and ten subposterior densities. When the number of subsets is increased to $M = 20$, the average estimated relative $L_2$ distance again increases for all four methods. The direct density product method has a reasonably low average across all parameters with a value of 0.039, with the remaining three methods having considerably larger average values, ranging from 0.283 to 1.030 (Table 1).

**Multivariate Normal with Exponential prior distribution**

Here, we illustrate the Multivariate Normal model with unknown mean vector $\boldsymbol{\mu}$ of dimension $d = 20$, and known variance-covariance matrix $\boldsymbol{\Sigma}$. The priors for the unknown mean parameters $\boldsymbol{\mu}$ are assigned non-conjugate Exponential distributions. We simulate data so that the posterior distributions for the individual $\mu_j$ parameters, $j = 1,...,20,$ will be non-Gaussian decreasing distributions, to further illustrate the performance of the four methods for non-Gaussian posteriors.

The data are simulated from a Multivariate Normal distribution, as follows:

$$\begin{aligned}
&y_r \sim \text{Normal}_d(\boldsymbol{\mu}, \boldsymbol{\Sigma}),\ r=1,\ldots,100{,}000;\ d=20;\\
&y_r = (y_{1r},\ldots,y_{20r})';\\
&\boldsymbol{\mu} = (\mu_1,\ldots,\mu_{20})';\\
&\boldsymbol{\Sigma}\ \text{is dimension } 20\times 20.
\end{aligned} \quad (19)$$

We use large means and large variances, with $\boldsymbol{\mu} = (1000,\ldots,1000)'$, and with the variance-covariance matrix $\boldsymbol{\Sigma}$ defined as follows:

$$\Sigma_{jj} = 1\times 10^8,\ j=1,\ldots,20;\ \Sigma_{jj'} = 2\times 10^7,\ j\neq j'. \quad (20)$$

The values for $\boldsymbol{\Sigma}$ were chosen so that the correlation among all parameters is $\rho = 0.2$.

Our Bayesian model is the following; we assigned non-conjugate Exponential priors to the individual $\mu_j$ parameters, $j=1,\ldots,20,$ so that the resulting posteriors are non-Gaussian decreasing distributions:

$$\begin{aligned}
&y_r \mid \boldsymbol{\mu} \sim \text{Normal}_d(\boldsymbol{\mu}, \boldsymbol{\Sigma}),\ r=1,\ldots,100{,}000;\ d=20;\\
&\boldsymbol{\mu}\ \text{unknown},\ \boldsymbol{\Sigma}\ \text{known},\\
&p(\mu_j) \sim \text{Exponential}(\lambda_j),\ j=1,\ldots,20;\ \lambda_j\ \text{known}.
\end{aligned} \quad (21)$$

Here, $\boldsymbol{\Sigma}$ is set equal to the simulated values specified above. The Exponential prior density for the full data model for each $\mu_j$ has the following form:

$$p(\mu_j \mid \lambda_j) \propto \exp(-\lambda_j \mu_j),\ j=1,\ldots,20. \quad (22)$$

For the full data model, we assign $\lambda_j = 1,\ j = 1,\ldots,20$. For the $M$ subset priors, the product is required to be proportional to the full data prior, as follows:

$$\prod_{m=1}^{M} \exp(-\lambda'_j \mu_j) \propto \exp(-\lambda_j \mu_j),\ j=1,\ldots,20. \quad (23)$$

This produces the following equivalence:

$$M \times \lambda'_j = \lambda_j, \ j = 1,...,20, \tag{24}$$

where $\lambda'_j$ are the subset prior parameters. Solving for $\lambda'_j$ forms the following:

$$\lambda'_j = \frac{\lambda_j}{M}, j = 1,...,20. \tag{25}$$

For $M = 10$ subsets, the resulting values are $\lambda'_j = 0.1$, and for $M = 20$ subsets, the resulting values are $\lambda'_j = 0.05$.

We generated posterior MCMC samples for each of the $\mu_j$ model parameters, $j = 1,...,20$, for the 10 data subsets as well as the full data set, and the results were non-Gaussian decreasing marginal posterior distributions for each of the $\mu_j$ model parameters. The direct density product method again resulted in the lowest average estimated relative $L_2$ distances for the marginal posterior distributions of the $\mu_j$, $j = 1,...,20$, with an average value of 0.0037. In contrast, the averages were considerably larger for the remaining three methods, with averages ranging from 1.081 to 1.330 (Table 1). The plots of the estimated combined posterior density and full data posterior density are shown in Figure 3 for the marginal of $\mu_1$ for all four methods; also shown are the full data posterior and ten subposterior densities. Note that the direct density product method produces skewed, non-Gaussian estimated marginal posterior distributions, similar to the full data posteriors, while the remaining three methods create Gaussian-shaped estimated posterior distributions for each of the $\mu_j$ parameters.

When the number of subsets is increased to $M = 20$, the average estimated relative $L_2$ distance again increases for all four methods. The direct density product method again has a reasonably low average across all parameters with a value of 0.0032, while the remaining three methods having considerably larger average values, ranging from 1.064 to 1.203 (Table 1).

**Computational time**

Next, we show the computational times for the four methods for each of the examples above (Table 2). Our direct density product method and the semiparametric DPE method both have lengthy computational times compared to the sample average and consensus Monte Carlo methods. For the model with only one unknown parameter, the semiparametric DPE method has the longest running time of all methods. For the multi-parameter models, our direct density product method has the longest running time. The lengthy computational time of our direct density product method is due to the density estimation for each subset; when the $x$-axis grid has a small interval width, the marginal density is estimated at a large number of values, which increases the computational time. The sample average method has the fastest computational times of the four methods for all models, and the consensus Monte Carlo method has the second fastest computational times.

**Discussion**

Here, we introduced the new direct density product method for parallel communication-free MCMC analyses that is particularly well suited to estimating marginal non-Gaussian posterior distributions. We found in simulation studies for three Bayesian models with non-Gaussian posteriors that our method outperformed three existing methods based on estimated relative $L_2$ distance.

Our new method, as well as the three existing methods used for comparison, are suitable for models that have unknown parameters with unchanging dimension in continuous parameter spaces. In addition, our method estimates only marginal posterior distributions and does not estimate joint distributions; this is similar to other Bayesian approaches such as INLA (Rue et al. [17]), where the focus is on estimation of posterior marginal distributions. Since our method estimates only marginal densities, it can be used for models with unlimited dimension, since it avoids the computational difficulties inherent in estimating joint posterior distributions in high dimensions. In contrast, the semiparametric DPE method can become computationally infeasible even for models with 50 unknown model parameters and MCMC sample size of 50,000. Another advantage of our method is that the densities can be approximated using any density estimation technique, while the semiparametric DPE is based on the Gaussian kernel density estimation method.

One limitation to our new method is that the approximation to the full data posterior distributions depends on the overlapping area of the subposterior distributions. When the subposterior distributions are far apart, the estimated posterior density has larger estimated relative $L_2$ distance. However, we assume the full data set has been partitioned randomly, and that the subposterior densities are independent, by the independent product equation in Equation (3); thus, the distance between subposterior densities is typically not expected to be extreme.

For Bayesian models that result in Gaussian posteriors for unknown model parameters, the consensus Monte Carlo method and the semiparametric DPE method will typically outperform our direct density product method and the sample average method. This is primarily due to the consensus Monte Carlo method and the semiparametric DPE method both taking into account the correlation among parameters to produce the joint posterior distributions. We thus only discuss and analyze non-Gaussian distributions here.

# Appendix

**Inverse cumulative distribution function method**

For our direct density product method above, a set of $S$ samples from each estimated full data marginal posterior density can be generated using the inverse cumulative distribution function sampling method. For this, repeat the following procedure $S$ times, where $S =$ the number of samples to generate; typically $S$ is the same as the number of MCMC samples $T$ (see also Devroye [32]):

a) Generate $U$ from the Uniform(0,1) distribution.

b) Compute and return $x = F^{-1}(U)$, where $F$ is the empirical cumulative distribution function created from the estimated full data marginal posterior density from our direct density product method.

The resulting set of $S$ values of $x$ has distribution $F$.

**Figure Legends**

**Figure 1.** Results for the simulation study of the Binomial data with conjugate beta prior, for the marginal of the $p$ parameter and $M = 10$ subsets. (a) 10 subposterior densities; (b) full data and estimated combined posterior densities for the four combining methods. The direct density product method produces the smallest $L_2$ distance (see Table 1).

**Figure 2.** Results for the simulation data of the Multinomial model with conjugate Dirichlet prior, for the marginal of the $p_1$ parameter and $M = 10$ subsets. (a) 10 subposterior densities; (b) full data and estimated combined posterior densities for the four combining methods. The direct density product method produces the smallest $L_2$ distance (see Table 1).

**Figure 3.** Results for the simulation data of the Multivariate Normal model with non-conjugate Exponential priors, for the marginal of the $\mu_1$ parameter and $M = 10$ subsets. (a) 10 subposterior densities; (b) full data and estimated combined posterior densities for the four combining methods. The direct density product method produces the smallest $L_2$ distance (see Table 1).

# Tables

**Table 1.** Average estimated relative $L_2$ distances

| Model | Number of Model Parameters, $d$ | Number of Subsets, $M$ | Direct Density Product Method | Consensus Monte Carlo Method | Semiparametric Density Product Estimator Method | Sample Average Method |
|---|---|---|---|---|---|---|
| Binomial with Beta Prior | 1 | 10 | 0.015 | 0.288 | 0.329 | 0.576 |
|  | 1 | 20 | 0.054 | 0.510 | 0.537 | 1.015 |
| Multinomial with Dirichlet Prior | 20 | 10 | 0.029 | 0.194 | 0.208 | 0.581 |
|  | 20 | 20 | 0.039 | 0.283 | 0.304 | 1.030 |
| Multivariate Normal with Exponential Priors | 20 | 10 | 0.0037 | 1.082 | 1.330 | 1.081 |
|  | 20 | 20 | 0.0032 | 1.064 | 1.203 | 1.064 |

Average estimated relative $L_2$ distances over all marginal posterior densities for the specified Bayesian models.

**Table 2.** Computational time (in seconds)

| Model | Number of Model Parameters, $d$ | Number of Subsets, $M$ | Direct Density Product Method | Consensus Monte Carlo Method | Semiparametric Density Product Estimator Method | Sample Average Method |
|---|---|---|---|---|---|---|
| Binomial with Beta Prior | 1 | 10 | 44.47 | 1.75 | 303.67 | 0.02 |
| | 1 | 20 | 92.88 | 3.30 | 630.08 | 0.03 |
| Multinomial with Dirichlet Prior | 20 | 10 | 556.63 | 4.63 | 427.11 | 0.18 |
| | 20 | 20 | 1,238.88 | 9.07 | 855.79 | 0.34 |
| Multivariate Normal with Exponential Priors | 20 | 10 | 819.85 | 7.11 | 690.39 | 0.17 |
| | 20 | 20 | 2,853.24 | 8.55 | 927.63 | 0.28 |

Computational times, in seconds, for the subset posterior MCMC samples combining methods. The results are based on a computer with operating system Windows 7 and an Intel Core i7-4600U CPU 2.1 GHz Processor.

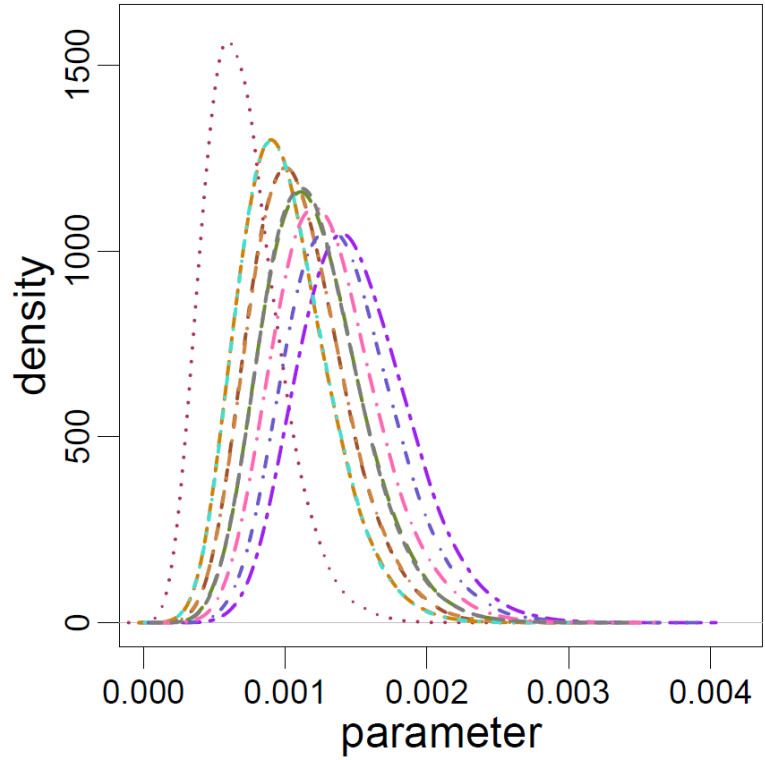

(a)

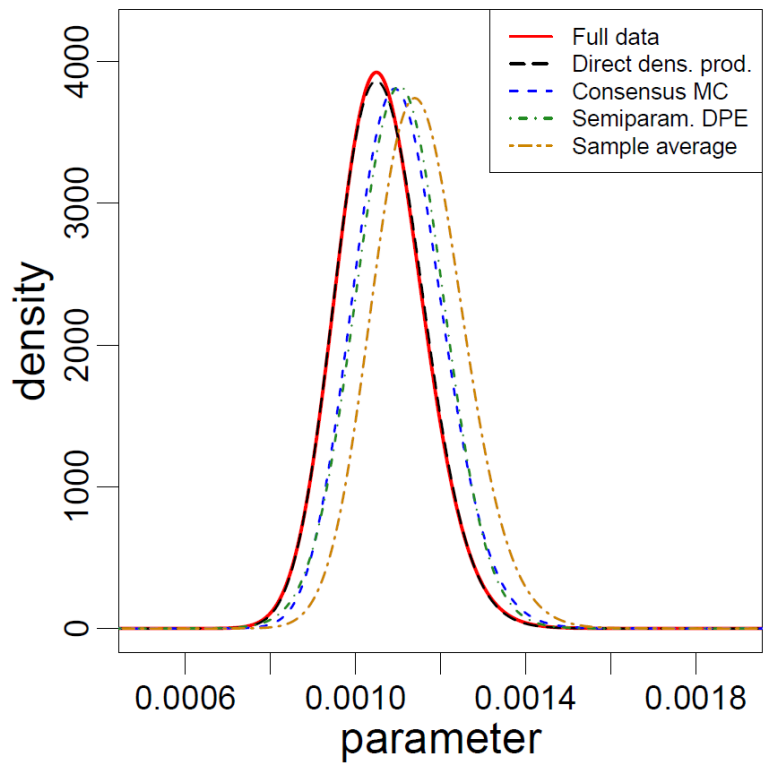

(b)

Figure 1

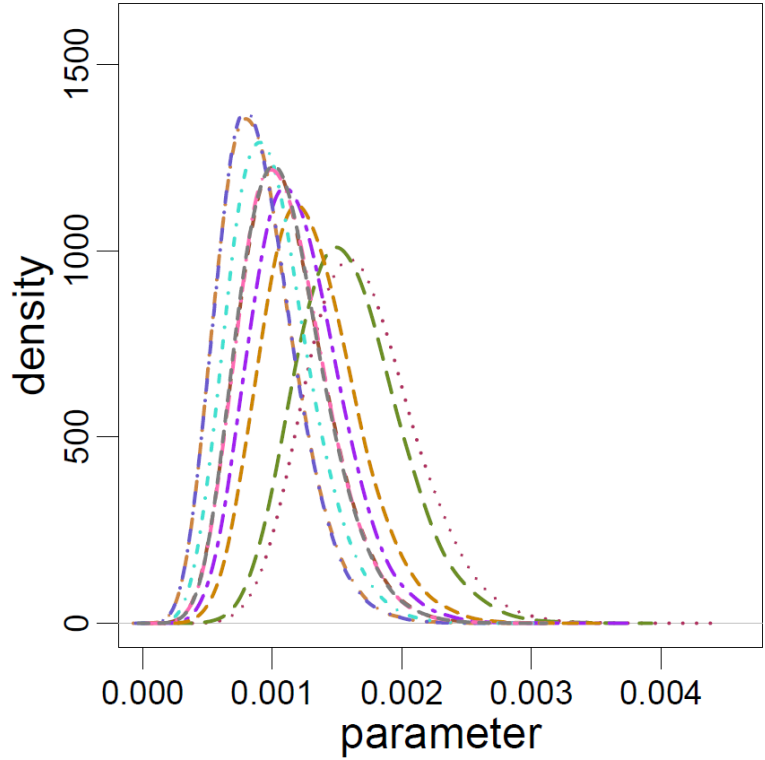

(a)

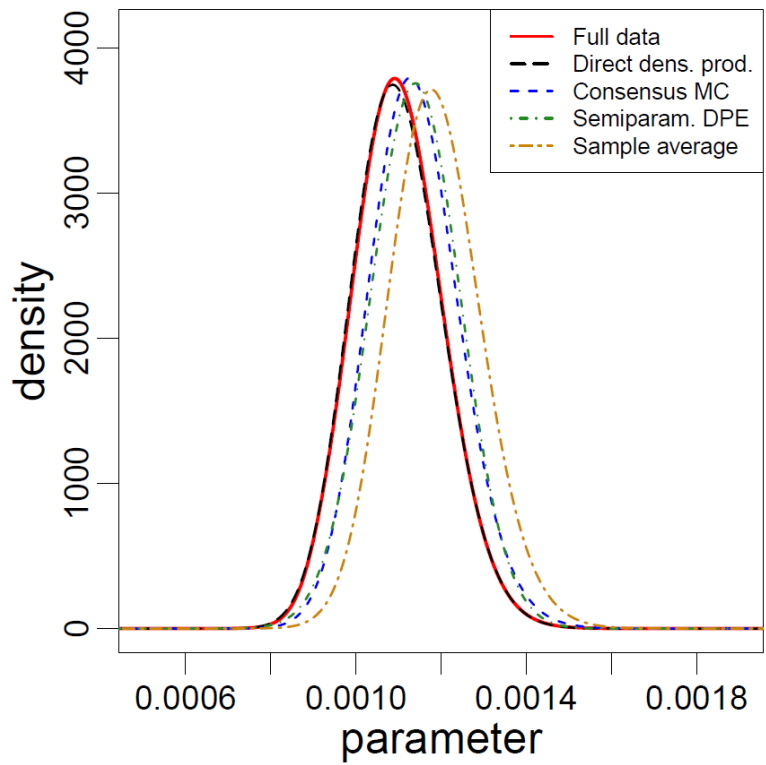

(b)

Figure 2

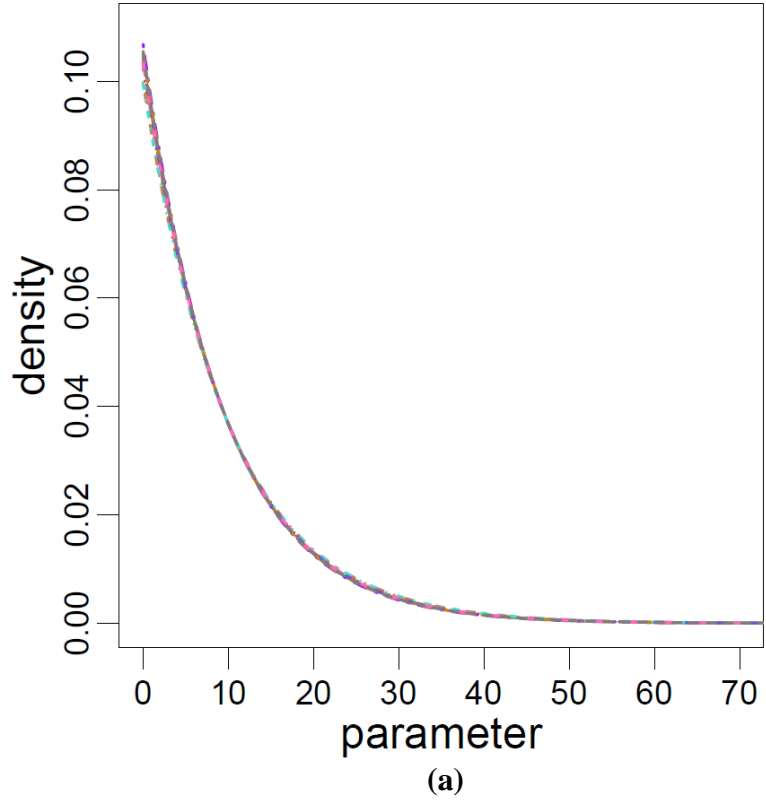

(a)

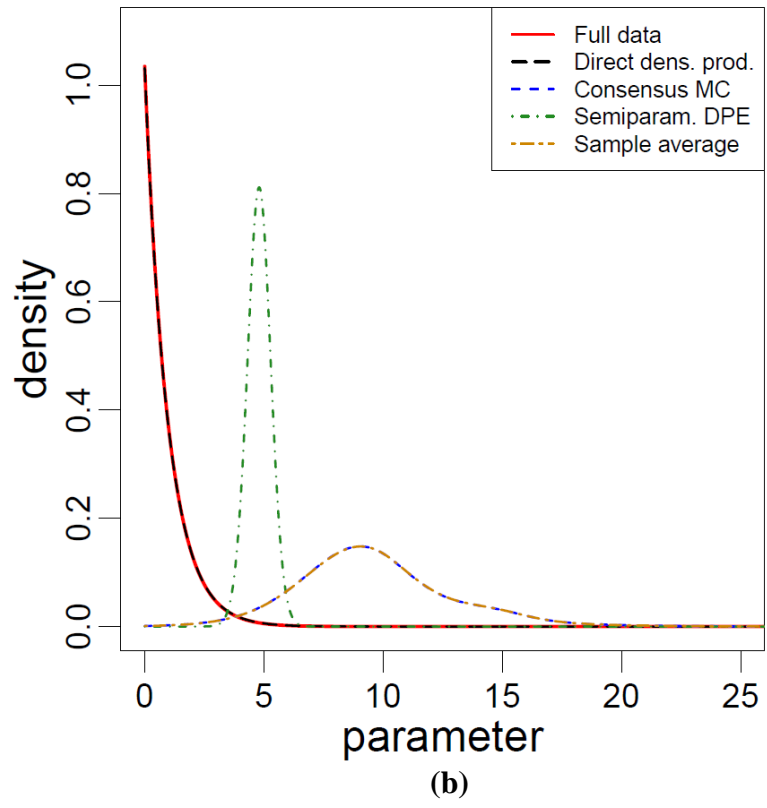

(b)

Figure 3